\documentstyle[12pt]{article}
\begin{document}
\title
{The $\zeta$-function answer to parity violation in three
dimensional gauge theories
}
\author{R.E.Gamboa Sarav\'\i\thanks{CONICET, Argentina}, G.L.Rossini$^*$
and F.A.Schaposnik\thanks{Investigador CICBA, Argentina}\\
{\normalsize \it Departamento de
F\'\i sica, Universidad Nacional de La Plata} \\
{\normalsize \it C.C. 67, (1900) La Plata, Argentina.}\\}
\date{}
\maketitle
\def\thepage{\protect\raisebox{0ex}{\ } La Plata 94-09}
\thispagestyle{headings}
\markright{\thepage}

\begin{abstract}
 We study parity violation in $2+1$-dimensional gauge theories
coupled to massive fermions. Using the $\zeta$-function regularization
approach we evaluate the ground state fermion current in an
arbitrary gauge field background, showing that it
gets two different contributions which violate parity invariance
and induce a Chern-Simons term in the gauge-field effective
action. One
is related to the well-known {\em classical} parity breaking produced by a
fermion mass
term in 3 dimensions; the other one,
already present for
massless fermions,  is related to peculiarities of gauge
invariant regularization in odd-dimensional spaces.
\end{abstract}
\newpage
\pagenumbering{arabic}

\section{Introduction}
Gauge theories in three dimensional space-time exhibit a variety
of phenomena of interest not only in Quantum Field Theory
\cite{Jac1} but also in Condensed Matter Physics \cite{Frad}.

An important feature of three dimensional kinematics concerns the
possibility of giving a (topological) mass to the vector field by
including an unconventional term in the gauge field Lagrangian
\cite{Jac2}-\cite{Scho}: the Chern-Simons (CS) secondary
characteristic.

The CS term, of topological origin, violates both P and T
invariances. Since the same happens for the mass term for a
(two-component) Dirac spinor in three dimensional space-time, it
is natural to expect an interesting interplay between both masses
in 3-dimensional gauge theories coupled to fermions. Indeed, in
refs.\cite{Jac2}-\cite{Jac3},\cite{NS}-\cite{Red} it was shown that if any of
the two
mass terms is
included in the Lagrangian, the other is then induced by
radiative corrections.

Now, in ref.\cite{NS}-\cite{Red}, it was also shown that even massless
fermions, when
coupled to gauge fields, generate a CS term.
Originally this effect was thought as a consequence of the
introduction of a fermion mass term within the Pauli Villars
regularization procedure \cite{Red}. However, the occurrence of
the CS term was confirmed in ref.\cite{GMSS} (hereafter referred
as I) using the $\zeta$-function approach, {\em where no regulating
mass term is added at any stage of the calculations}.  In fact,
the violation of parity in odd
dimensions  (and, consequently, the generation of a CS term)
is
analogous to the non-conservation of the axial current in even
dimensions:
the imposition of gauge invariance produces in both cases an
anomaly for a
symmetry (parity, chiral symmetry) of the original action for
{\it any}
sensible regularization.

The issue of parity violation in three dimensions is  relevant in
different contexts. In particular, discrepancies on overall
parity breaking have arisen in the
analysis of the $2+1$ Thirring model \cite{HP}-\cite{Gom}, this
leading to contradictory results concerning dynamical mass
generation for fermions (see \cite{RoS} for a treatment of the Thirring model
using the techniques described in this paper). Since
$3+1$ dimensional field theories become effectively
$3$-dimensional in the high
temperature limit, the problem has applications also in $4$
dimensional Physics.
Also in Condensed Matter Physics, the appearence of Chern-Simons
term through parity anomaly provides an adequate ground for
testing the physics of anyons in three dimensional fermionic
systems \cite{Frad}.  In this respect, generation of CS
term through the fermion effective action has direct consequences
on the thermodynamic properties of the system \cite{Nino}.

\vspace{0.5 cm}
Since the generation of the CS term is a consequence of
regularizing the three dimensional fermionic ground state
current, a careful analysis of the regularization
prescription is needed in order to decide whether or not parity
violation occurs.
To this issue we address in the present work, extending previous
results that we
have  already presented in I.

Firstly, we give a mathematically
rigourous scheme leading to the definition of the path-integral
fermionic measure for a $3$-dimensional gauge theory.
This is achieved through  a careful treatment of the fermionic
determinant
(a not well defined object for the unbounded Dirac operator)
using the results of Seeley \cite{Seeley} on complex powers of
elliptic operators.

Secondly, we give
the recipe for
computing fermionic ground state currents. Since at this stage we
already dispose
of a finite expression for  the fermionic partition function ${\cal{Z}}$
within the $\zeta$-function approach, this recipe reduces to give
explicit formul{\ae} for obtaining vacuum expectation values by
using the differentiability properties of the $\zeta$-function
\cite{GMS}.

Finally, we apply our approach to the analysis of
massive fermions
thus completing the work initiated in I on
massless fermions.
We think that the present analysis clarifies the origin of the
parity anomaly and, in particular, the appearence of two contributions to
parity violation, one
originated in peculiarities of calculations in odd-dimensional
spaces,
the other arising from the parity violation produced, already at the
classical level, by the fermion mass
in 3 dimensions.
\vspace{0.5 cm}

In respect with massless fermions, the original calculation of
the ground state fermion current $j_{\mu}[A]$ in a constant
magnetic background $ {^* \! F}_{\mu}$,
$ ^* \! F_{\mu} = i\epsilon_{\mu \nu \alpha}\partial_{\nu}
A_{\alpha}$ was presented in ref.\cite{Red} (we are working here in Euclidean
space). As an example, the
perturbative calculation using Pauli-Villars regularization
yields for the Abelian case, in the one loop approximation
\cite{Red},

\begin{equation}
j_{\mu}[A] = \frac{m}{\vert m \vert} \frac{
e}{4\pi}{^*\!F_{\mu}}
\label{redlich}
\end{equation}
where $e$ is the gauge coupling constant and $m$ is a fermionic
mass to be put to zero at the end of the calculations. This current is
identically conserved, but it violates parity conservation
since $^*\!F_{\mu}$ is a pseudovector.

{}From this current the effective action $\Gamma[A]$ can be
computed,
using the relation
\begin{equation}
\frac{\delta \Gamma[A]}{\delta A_{\mu}} = e j_{\mu}[A].
\label{accion}
\end{equation}
{}From eqs.(\ref{redlich})-(\ref{accion}) we see that the effective
action, appart from parity conserving contributions, contains a
CS term,

\begin{equation}
\Gamma[A]_{Pauli-Villars} =  \frac{m}{\vert m \vert} \frac{e^2}{2}
S_{CS}[A] + S_{PC}[A]
\label{CS}
\end{equation}
where  $S_{CS}[A]$ is the (Abelian) Chern-Simons action,
\begin{equation}
S_{CS}[A] = \frac{1}{4\pi}\int d^3x A_{\mu}{^*\!F_{\mu}}
\label{CS1}
\end{equation}
and $S_{PC}[A]$ are parity conserving terms. We have explicitly
indicated in $\Gamma$ that
this result was obtained in \cite{Red} within the Pauli-Villars
regularization scheme.

Of course, this result can be trivially generalized to the case
of $N$ flavors.
Instead of (\ref{CS}), one has when there are $N$ fermion
species:
\begin{equation}
\Gamma[A]_{Pauli-Villars} =  \sum_{i=1}^{N}\frac{m_i}{\vert m_i
\vert} \frac{e^2}{2} S_{CS}[A] + S_{PC}[A,N].
\label{CSN}
\end{equation}
As in the $N=1$ case, one has to put the fermion masses $m_i = 0$
to recover the massless case. Within this approach,
one could evidently choose the signs of $m_i$'s in such a way
that,
for even $N$, the overall CS contribution to $\Gamma$ cancells
out so that overall parity would not be violated. Now, using the
$\zeta$-function approach,
we have proven in I, without necessity of introducing a
mass-parameter,  that
instead of (\ref{CS}) one directly obtains:
\begin{equation}
\Gamma[A]_{\zeta} =  \pm \frac{e^2 }{2} S_{CS}[A] + S_{PC}[A]
\label{CSnos}
\end{equation}
As it will become clear in the next section,  within the
$\zeta$-function approach, the sign ambiguity in (\ref{CSnos})
can be traced back to the choice of an integration path $\Omega$
in the complex plane, necessary for defining the complex powers
of the Dirac operator. In odd-dimensional space-times, the choice
of $\Omega$ in the upper (lower) half
plane yields to a positive (negative) sign in (\ref{CSnos}).
Then, once a
definite integration path is chosen, consistency in the
definition of complex powers implies that, when $N$ species are
present,  all fermions contribute
with the same sign so that, instead of eq.(\ref{CSN}) one gets

\begin{equation}
\Gamma[A]_{\zeta} =  \pm  N \frac{e^2}{2} S_{CS}[A] + S_{PC}[A,N]
\label{CSnosN}
\end{equation}
and overall parity is always violated.

To end with the resum\'e of the massless case analysis, it should be stressed
that the $\zeta$-function approach allows the
calculation of the parity violating contribution to $j_{\mu}[A]$
for
{\it arbitrary} $A_{\mu}$ (and not necessarily for one leading to
a constant magnetic field) in an {\it exact} form.
As explained above, it was shown in Ref.\cite{Red} that
the parity violation contribution was induced from one fermion
loops and arguments were given to discard higher loop
corrections. Now, we have proven in I that the Chern-Simons
contribution present in (\ref{CSnos}) is {\it all}
the parity violating contribution one has to expect for three
dimensional massless fermions since, as in the case of the chiral anomaly,
the $\zeta$-function
approach gives the exact anomalous contribution.

Let us come now to the massive fermion case. The problem was
already discussed in \cite{Jac2}, to the one-loop order and in a
constant magnetic background, using
Pauli-Villars regularization and clearly explained in
\cite{Jac1}.
The conclusion was that when massive fermions couple to the gauge
field, a CS
term  is induced by fermion radiative corrections. The fact that
a fermion mass term in three dimensions violates P and T was
considered at the origin of this result.
The explicit form for the effective action in the massive case
was calculated in \cite{Red}. The result coincides
with (\ref{CS}) except that in this case $m$ is the physical
fermion mass.
Again, the conclusion was dependent on how to regularize
divergent objects \cite{Jac2},\cite{KaSu}-\cite{BeLe}.
It is then worthwhile to analyse the problem of massive fermions
using the
$\zeta$-function approach developed in I. We here undertake such
program
explaining in section 2 the mathematical tools needed for the
computation
of ground state fermion currents and leaving for section 3 the
explicit
calculation of $j_{\mu}$. A summary of results and discussions is
given in Section 4.

\section{How to compute currents}

We first consider the Abelian case giving the basic formul{\ae}
to be used in the next section.

The partition function for a massive Dirac fermion doublet in $d$=3
Euclidean space-time dimensions is:
\begin{equation}
{\cal{Z}}[A_{\mu}] = \int {\cal{D}}\bar\psi {\cal{D}}\psi
\exp[-\int \bar\psi D[A] \psi d^3x]
\label{1}
\end{equation}
where $A_{\mu}$ is a background vector field and the Dirac
operator $D[A]$ is given by\footnote{Our conventions for $\gamma$
matrices are $\gamma_{\mu} = \sigma_{\mu}$ with $\sigma_{\mu}$
($\mu=1,2,3$) the Pauli matrices.}
\begin{equation}
D[A] = \gamma_{\mu}(i\partial_{\mu} + eA_{\mu})+ im.
\label{2}
\end{equation}
In (\ref{1}), ${\cal{D}}\bar\psi {\cal{D}}\psi$ is some fermionic measure to be
defined below. The results we shall describe are valid for any
elliptic operator $L$, not necessarily Hermitian, defined on a
compact manifold without boundary \cite{GMS},\cite{Anna}.
We shall assume however that
they are also valid for $R^3$.

The ground state current $j_{\mu}[A]$ in the presence of the
background field,
\begin{equation}
j_{\mu}[A] =\langle \bar\psi(x)\gamma_{\mu}\psi(x)\rangle_A,
\label{3}
\end{equation}
can be calculated from ${\cal{Z}}[A]$ as
\begin{equation}
j_{\mu}[A]=-\frac{1}{e}\frac{\delta}{\delta A_{\mu}(x)}\log {\cal{Z}}[A].
\label{4}
\end{equation}
Of course, since the Dirac operator is unbounded, ${\cal{Z}}[A]$ needs a
regularization,
\begin{equation}
{\cal{Z}}_{reg}[A] \equiv
\mbox{det}D[A]\left|\rule[-0.4cm]{0cm}{.4cm}_{reg}\right. .
\label{5}
\end{equation}
We shall adopt in this work the $\zeta$-function regularization
method which automatically ensures {\em local gauge invariance}. We
thus define as usual
\begin{equation}
\zeta(D[A],s)=\mbox{Tr}(D^{-s}[A]),
\label{6}
\end{equation}
since $D[A]$ is invertible. From (\ref{6}) we can define
\begin{equation}
{\cal{Z}}_{reg}[A]=\exp(-\frac{d}{ds}\zeta(D[A],s))\left|\rule[-0.4cm]{0cm}{.4cm}_{s=0}\right.
\label{7}
\end{equation}
where the right-hand side should be computed for complex $s$
(with large enough real part) and then analytically extended to
$s=0$. This definition amounts in
practice to the definition of the fermionic path-integral
measure.
It has been proven in \cite{GMS} that ${\cal{Z}}[A]$ written as in
eq.(\ref{7}) is differentiable; then, one can easily obtain from
eqs.(\ref{4}),(\ref{7}) a regularized expression for
$j_{\mu}[A]$. Indeed, one can write
\begin{equation}
 \int\frac {\delta}{\delta A_{\mu}(x)}\log {\cal{Z}}_{reg}[A] a_{\mu}(x)
d^3x =
\frac{d}{dt}\log{\cal{Z}}_{reg}[A_{\mu}+ta_{\mu}]\left|\rule[-0.4cm]{0cm}{.4cm}_{t=0}\right. .
\label{8}
\end{equation}
(Here $a_{\mu}$ indicates a direction in $A_{\mu}$ functional
space along which the derivative is taken). Now, from eq.(\ref{7}) one
has
\begin{equation}
\int\frac {\delta}{\delta A_{\mu}(x)}\log {\cal{Z}}_{reg}[A] a_{\mu}(x)
d^3x=
\frac{d}{dt}\left(-\frac{d}{ds}\zeta(D[A_{\mu}+ta_{\mu}],s)\left|\rule[-0.4cm]{0cm}{.4cm}_{s=0}\right.
\right)\left|\rule[-0.4cm]{0cm}{.4cm}_{t=0}\right.
\label{9}
\end{equation}
or, using eq.(\ref{6}) and interchanging the order of the derivatives,
\begin{equation}
\int\frac {\delta}{\delta A_{\mu}(x)}\log {\cal{Z}}_{reg}[A] a_{\mu}(x)
d^3x=
\frac{d}{ds}\left[ s \mbox{Tr} ( D^{-s-1}[A]
e\gamma_{\mu}a_{\mu})\right]\left|\rule[-0.4cm]{0cm}{.4cm}_{s=0}\right. .
\label{10}
\end{equation}
In eqs.(\ref{6}),(\ref{10}), ``$\mbox{Tr}$'' stands for the operator
trace, which includes the matrix trace ``$\mbox{tr}$'' and an
integration over the space-time of the diagonal of the operator
kernel. If we denote by $K_s(x,y;D[A])$ the kernel of the
operator $D^s[A]$,
\begin{equation}
D^s[A]f(x)=\int d^3y K_s(x,y;D[A]) f(y)
\label{kernel}
\end{equation}
we can read from eq.(\ref{10})
\begin{equation}
\frac {\delta}{\delta A_{\mu}(x)}\log {\cal{Z}}_{reg}[A]=e\frac{d}{ds}\left\{
s\, \mbox{Tr}
\left[\gamma_{\mu}
K_{-s-1}(x,x;D[A])
 \right]\right\} \left|\rule[-0.4cm]{0cm}{.4cm}_{s=0}\right. .
\label{derivative}
\end{equation}
Then, from eqs.(\ref{4})-(\ref{derivative}), one has
\begin{equation}
{j_{\mu}}^{reg}[A]= -\frac{d}{ds}\left\{ s\, \mbox{Tr}
\left[\gamma_{\mu}
K_{-s-1}(x,x;D[A])
 \right]\right\} \left|\rule[-0.4cm]{0cm}{.4cm}_{s=0}\right. .
\label{11}
\end{equation}
This is the key formula we shall employ to compute fermionic currents in the
$\zeta$-function regularization scheme.

Note that $ K_{-s-1}(x,y;D[A])|_{s=0}$ is nothing but the Green
function $G(x,y)$ of the operator $D[A]$ when $x\neq y$.
Correspondingly, it is singular on the diagonal $x=y$.
Now,  for any elliptic operator $L$ of positive order $r$ in a space of
dimension $n$,
the kernel of $L^s$ is a
continuous function of $x,y$ for $Re(s)<-n/m$ that admits an analytic
extension to
the whole complex plane $s$ if $x\neq y$. On the diagonal $x=y$ it
is a meromorphic function of $s$ that has at
most simple poles at $s=(-n+j)/r$, $j=0,1, \dots$ \cite{Seeley}. For the Dirac
operator
$D[A]$ in three-dimensional Euclidean space-time, $ K_{-s-1}(x,x;D[A])$ has a
simple pole at $s=0$.
It is then clear that eq.(\ref{11}) gives ${j_{\mu}}^{reg}$ as the finite part
of the
Laurent series for
$-tr [\gamma_{\mu} K_{-s-1}(x,x;D[A])]$ at $s=0$, thus providing an
operative (finite) formula for $-tr[\gamma_{\mu}G(x,x)]$
\begin{equation}
-\frac{d}{ds}\left\{ s\, \mbox{tr}
\left[\gamma_{\mu} K_{-s-1}(x,x;D[A])
\right]\right\}\left|\rule[-0.4cm]{0cm}{.4cm}_{s=0}\right.
=-tr[ \gamma_{\mu}G^{reg}(x,x)].
\label{12}
\end{equation}
The right hand side in (\ref{12}) gives the $\zeta$-function analogous of the
usual
expression that can be found in the literature \cite{Schwinger} for
the ground state current, with the
regularization procedure specified in the present approach by the left hand
side.

We see at this point that for computing ${j_{\mu}}^{reg}[A]$ we need an
explicit formula for the
kernel $K_{-s-1}(x,y;D[A])$ so as to evaluate the r.h.s.\ of
eq.(\ref{11}). For the sake of clarity and consistency, let us present
some of the results of Seeley \cite{Seeley} on complex powers of
elliptic operators in the restricted form  needed for studying the Dirac
operator.

The complex powers of the elliptic differential operator $D[A]$ (of order 1 in
a space of dimension
$n$=3) are best described
in terms of the symbol of the operator and the symbol of its
resolvent $(D[A]-\lambda)^{-1}$. The symbol of
$D[A]$	is a polynomial $\sigma(D[A])$ in a vector $\xi_{\mu}$ that can
be thought as the Fourier
variable associated to $x_{\mu}$. It is obtained from $D[A]$ by replacing
$-i\partial_{\mu} \to \xi_{\mu}$ and takes the form
\begin{equation}
\sigma(D[A])(x,\xi)=-\gamma_{\mu}\xi_{\mu}+e\gamma_{\mu}A_{\mu}(x)+im
\equiv	 a_{1}(x,\xi)+a_{0}(x,\xi)
\label{13}
\end{equation}
where $a_j(x,\xi)$ ($j$=0,1) are homogeneous functions of degree $j$
in $\xi_{\mu}$.

The resolvent $(D[A]-\lambda)^{-1}$ of the differential operator $D[A]$ is a
pseudo-dif\-feren\-tial operator
\cite{Hormander}. Its symbol is a generalization of the definition above
and can be properly
approximated by
\begin{equation}
\sigma((D[A]-\lambda)^{-1})(x,\xi)=\sum_{j=0}^{\infty} b_{-1-j}(x,\xi;\lambda).
\label{15}
\end{equation}
Here $b_{-1-j}(x,\xi;\lambda)$ are homogeneous functions of degree
$-1-j$ in $\xi_{\mu}$ and $\lambda$. These coefficients can be
recursively evaluated for
each order in $\xi_{\mu}$ from the relation\footnote{The product of
symbols is
defined in such a way that that it corresponds to the symbol of the
composition of
operators \cite{Hormander}.}
\begin{equation}
\sigma((D[A]-\lambda)^{-1})  \circ \sigma((D[A]-\lambda))=I
\label{16}
\end{equation}
provided one takes $\lambda$ order $\xi$, so that it
combines
with the top term $a_1$ of the symbol of $D[A]$. The recursive formula reads
\begin{equation}
b_{-1}(a_1-\lambda)=I,
\label{17}
\end{equation}
\[
b_{-1-l}(a_1-\lambda)+\sum_{j=0}^{l-1}\sum_{\alpha=0}^{l-j}\frac{1}{\alpha !}
(\frac{\partial}{\partial\xi})^{\vec{\alpha}}b_{-1-j}(-i\frac{\partial}
{\partial x})^{\vec{\alpha}}a_{1-l+\alpha+j}=0, \mbox{~~~~$l$=1,2,{\cal{D}}ots}
\]
where $\vec{\alpha}=(\alpha_1,\alpha_2,\alpha_3)$  is a vector of non
negative integers, $\alpha=\sum_{i=1}^{3}\alpha_i$ and
$(\frac{\partial}{\partial\xi})^{\vec{\alpha}}= \prod_{\mu=1}^{3}
(\frac{\partial}{\partial\xi_{\mu}})^{\alpha_{\mu}}$, $
(\frac{\partial}{\partial x})^{\vec{\alpha}}=
\prod_{\mu=1}^{3}(\frac{\partial}{\partial x_{\mu}})^{\alpha_{\mu}}$ .

{}From eqs.(\ref{13}),(\ref{17}) the coefficients of the symbol of the
resolvent can be evaluated for the Dirac operator. The first three of them are
\begin{equation}
b_{-1}(x,\xi;\lambda)=\frac{1}{\lambda^2-\xi^2}(\not\! \xi -\lambda),
\label{18}
\end{equation}
\begin{equation}
b_{-2}(x,\xi;\lambda)=-\frac{1}{(\lambda^2-\xi^2)^2}(\not\! \xi
-\lambda)(e\not\!\! A+im)
(\not\! \xi-\lambda),
\label{19}
\end{equation}
\begin{eqnarray}
\lefteqn{b_{-3}(x,\xi;\lambda)=\frac{1}{(\lambda^2-\xi^2)^3}(\not\! \xi
-\lambda)
(e\not\!\! A+im)(\not\! \xi-\lambda)
(e\not\!\! A+im)(\not\! \xi-\lambda)+ }  \nonumber\\
& & ie\frac{\partial_{\mu}A_{\nu}}{(\lambda^2-\xi^2)^3}
(2\xi_{\mu}(\not\! \xi-\lambda)\gamma_{\nu}(\not\! \xi
-\lambda)+(\lambda^2-\xi^2
)\gamma_{\mu}\gamma_{\nu}(\not\! \xi -\lambda)).
\label{20}
\end{eqnarray}

Following Seeley \cite{Seeley}, the complex powers $D^s[A]$ can be defined as a
generalization of
Cauchy integral representation of $z^s$,
\begin{equation}
z^s=\frac{i}{2\pi}\oint_C \frac{\lambda^s}{z-\lambda}d\lambda.
\label{29}
\end{equation}
Here, the contour $C$ must avoid a log-like cut in the complex
plane $\lambda$  and encircle
the pole of the integrand.  For the operator $D[A]$, one must take
a contour  $\Omega$ that encircles
the whole spectrum of  $D[A]$ and avoids a ray going from the
origin to  infinity in a direction such
that no eigenvalue of $a_1$ lies on it (such a ray is called a
ray of	minimal growth or Agmon ray). For $Re(s)<0$
one can take $\Omega $ as a curve  begining at $\infty$, passing
along the  ray of minimal growth to
a small circle about the origin, then clockwise around the
circle, and  back to $\infty$ along the ray.
 The ray along which curve $\Omega$ is defined
cannot coincide with the real axis, since being
$D[A]$ Hermitian its eigenvalues are real.
Two possible $\Omega$ curves are illustrated in Fig.\ 1.

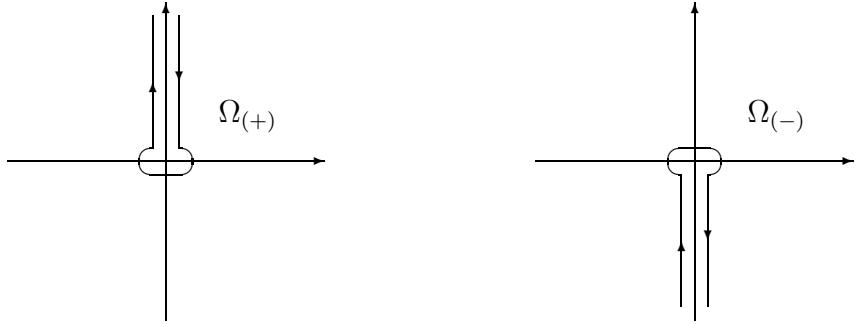
\begin{figure}
\begin{center}
\begin{picture}(390,200)(-200,-100)
\put(-160,0){\vector(1,0){120}}
\put(-100,-60){\vector(0,1){120}}

\put(-95,55){\vector(0,-1){25}}
\put(-95,30){\line(0,-1){25}}

\put(-105,5){\vector(0,1){25}}
\put(-105,30){\line(0,1){25}}

\put(-95,0){\oval(10,10)[r]}
\put(-105,0){\oval(10,10)[l]}
\put(-105,-5){\line(1,0){10}}

\put(-80,15){$\Omega_{(+)}$}

\put(40,0){\vector(1,0){120} }
\put(100,-60){\vector(0,1){120}}

\put(95,-55){\vector(0,1){25}}
\put(95,-30){\line(0,1){25}}

\put(105,-5){\vector(0,-1){25}}
\put(105,-30){\line(0,-1){25}}

\put(95,0){\oval(10,10)[l]}
\put(105,0){\oval(10,10)[r]}
\put(95,5){\line(1,0){10}}

\put(120,15){$\Omega_{(-)}$}

\end{picture}

\end{center}

\caption{Contours $\Omega$ for defining complex powers of $D[A]$.}
\label{figura1}
\end{figure}

Then, for $Re(s)<0$, one defines
\begin{equation}
D^s[A]=\frac{i}{2\pi}\int_{\Omega} \lambda^s (D[A]-\lambda)^{-1}d\lambda.
\label{30}
\end{equation}
This definition is analytically extended to the whole complex
plane through multiplication by integer powers of $D[A]$,
\begin{equation}
D^s[A]=D^k[A] D^{s-k}[A] \mbox{~~~~~~~~~$k$ integer, ~~$-1\leq Re(s)-k<0.$}
\label{30ext}
\end{equation}
The corresponding symbol is constructed using eq.(\ref{15}) and reads
\begin{equation}
\sigma(D^s[A])(x,\xi)=\sum_{j=0}^{\infty} \frac{i}{2\pi}\int_{\Omega} \lambda^s
b_{-1-j}(x,\xi;\lambda)d\lambda \equiv
\sum_{j=0}^{\infty} C_j(x,\xi;s),
\label{31}
\end{equation}
where
\begin{equation}
C_j(x,\xi;s)=\frac{i}{2\pi} \int_{\Omega} \lambda^s b_{-1-j}(x,\xi;\lambda)
d\lambda
\label{32}
\end{equation}
are homogeneous functions of (complex) degree $s-j$.
It is worthwhile noting that, while eq.(\ref{30}) is valid
for $Re(s)<0$, the relation between symbols eq.(\ref{31}) is valid for any
value of
$s$.

Using the tools described above we are now in conditions to write the
kernel $K_{-s-1}(x,\xi;D[A])$ for the operator $D^{-s-1}[A]$, necessary for
evaluating the ground
state current (eq.(\ref{11})).
Indeed, one of the main
points in Seeley's work \cite{Seeley} is that one can approximate kernel $K_s$
through Fourier transforms of
the $C_j$ coefficients,
\begin{equation}
K_s(x,y;D[A])=\sum_{j=0}^{N}\int \frac{d\xi}{(2\pi)^n}
C_j(x,\xi;s)e^{i(x-y).\xi}
+ R(x,y;s).
\label{33}
\end{equation}
Here $R(x,y;s)$ includes terms which will not contribute to the derivative in
eq.(\ref{11}) at
$s=0$ if $N$ is taken sufficiently large.
Though the integral in each term
converges only for $Re(s)< (-n+j)$ (being $n$ the dimension of spacetime,
here $n$=3), this expression can be
analytically extended to arbitrary $s$ for $x\neq y$. On the diagonal $x=y$,
$K_s(x,x;D[A])$ is extended to a meromorphic function with at most simple
poles at $s=(-n+k)$, $k=0,1, \dots$, one arising from each term included
in the sum in eq.(\ref{33}).

In order to compute  ${j_{\mu}}^{reg}$ in eq.(\ref{11}) one uses eq.(\ref{33})
to
evaluate the finite part of the simple pole of
$K_s(x,x;D[A])$ at $s=-1$. To this end, we shall use the following proposition
(proven in I and adapted here to the case of the Dirac operator):

\vspace{5mm}
{\em Proposition:} For the elliptic invertible first order operator $D[A]$ on a
3-dimensional compact manifold $M$ without boundary, the following identity
holds:
\begin{eqnarray}
\lefteqn{\frac{d}{ds}(sK_{s-1}(x,x;D[A]))\left|\rule[-0.4cm]{0cm}{.4cm}_{s=0}\right.
=}\nonumber\\
& &\lim_{y\to x} \left\{ G(x,y)-
\int \frac{d^3\xi}{(2\pi)^3}C_1(x,\xi / |\xi|;-1)|\xi|^{-2}e^{i\xi
.(x-y)} \right. \nonumber\\
& &\left. - \! \int \!\frac{d^3\xi}{(2\pi)^3}
C_0(x,\xi / |\xi|;-1)|\xi|^{-1}e^{i\xi .(x-y)}- \!
 \int_{|\xi|\geq 1} \! \frac{d^3\xi}{(2\pi)^3}C_2(x,\xi;-1)e^{i\xi
.(x-y)}\right\} \nonumber \\
& & -\int_{|\xi|=1} \frac{d}{ds}C_2(x,\xi;s)
\left|\rule[-0.4cm]{0cm}{.4cm}_{s=-1}\right. \frac{d^2\xi}{(2\pi)^3}
\label{proposition}
\end{eqnarray}
where $G(x,y)$ is the Green function of $D[A]$. (The interested reader can find
the
proof in I.)

\vspace{5mm}
The last term between brackets in eq.(\ref{proposition}) can be conveniently
rewritten as
\cite{Calderon} (see I)
\begin{eqnarray}
\lefteqn{\int_{|\xi|\geq 1} \frac{d^3\xi}{(2\pi)^3}C_2(x,\xi;-1)e^{i\xi
.(x-y)}=}
\nonumber \\
& & h_0(x,x-y)+M(x)( \log|x-y|+C(x-y) )
\label{h0MC}
\end{eqnarray}
where $h_0(x,z)$ is a homogeneous function of degree zero defined as
\begin{equation}
h_0(x,z)=\int \frac{d^3\xi}{(2\pi)^3}\mbox{P.V.} [C_2(x,\xi /|\xi|;-1)
-M(x)]|\xi|^{-3}e^{i\xi .z},
\label{h0}
\end{equation}
$M(x)$ is defined as
\begin{equation}
M(x)=\frac{1}{\omega}\int_{|\xi|=1} C_2(x,\xi;-1)d^2\xi
\label{M}
\end{equation}
and finally
\begin{equation}
\omega(\log|z|+C(z))=-\int_{|\xi|\geq 1}  \frac{d^3\xi}{(2\pi)^3}|\xi|^{-3}
e^{-i\xi .z}
\label{omega}
\end{equation}
where $C(z)$ is a regular function in the neighborhood of $z=0$. All Fourier
transforms are taken in the sense of distributions and P.V.\ means principal
value.

The interpretation of eq.(\ref{proposition}) is as follows:
The last three terms between brackets substract the singularities of $G(x,y)$
on the diagonal $x=y$.
Concerning the last term in eq.(\ref{proposition}), it is a {\em local} regular
term generated through the
regularization procedure that adds to the regular part of $G(x,x)$.
In view of this
interpretation we rewrite the result in eq.(\ref{proposition}) in the form
\begin{equation}
\frac{d}{ds}(sK_{s-1}(x,x;D[A]))\left|\rule[-0.4cm]{0cm}{.4cm}_{s=0}\right.
=G^{reg}(x,x)=G^{(subst)}(x) + G^{(local)}(x)
\label{last1}
\end{equation}
where
\begin{eqnarray}
\lefteqn{G^{(subst)}(x)=\lim_{y\to x} \left\{ G(x,y)-
\int \frac{d^3\xi}{(2\pi)^3}C_1(x,\xi / |\xi|;-1)|\xi|^{-2}e^{i\xi
.(x-y)} \right.}\label{last2} \\
& &- \! \int \!\frac{d^3\xi}{(2\pi)^3}C_0(x,\xi / |\xi|;-1)|\xi|^{-1}e^{i\xi
.(x-y)}
-
\left. \! \int_{|\xi|\geq 1} \!\frac{d^3\xi}{(2\pi)^3}C_2(x,\xi;-1)e^{i\xi
.(x-y)}\right\} \nonumber
\end{eqnarray}
and
\begin{equation}
G^{(local)}(x)= -\int_{|\xi|=1} \frac{d}{ds}C_2(x,\xi;s)
\left|\rule[-0.4cm]{0cm}{.4cm}_{s=-1}\right. \frac{d^2\xi}{(2\pi)^3}
\label{last3}
\end{equation}

In spite of its apparent complexity, the result in eq.(\ref{proposition}) can
be used with great
simplicity for evaluating ground state currents {\em for arbitrary gauge
background fields}.
This will be seen in the next section, where we shall employ
eqs.(\ref{11}),(\ref{proposition}) to compute the parity
violating terms of ${j_{\mu}}^{reg}[A]$.

\section{The current for massive fermions}

Formula (\ref{proposition}) is at the root of the $ \zeta $ \-function
 regularization prescription for
the calculation of ground state fermion currents.
 Using this approach, parity violating contribution to the ground
 state current was evaluated in I for
the case of massless fermions. In this section the analysis is
extended to the case of massive fermions
 with partition function given	by
 eq.(\ref{1}).

Let us start by understanding why parity violating terms should
 be expected in
a 2+1 fermionic theory. First,
note that a 3-dimensional fermion mass term constructed from
two-component spinors violates parity and time-inversion. Indeed,
 under parity transformation the vector and fermion fields behave as

\begin{eqnarray}
{\cal{P}} A_0(x) {\cal{P}}^{-1} &=&  A_0(x') \nonumber\\
{\cal{P}} A_1(x) {\cal{P}}^{-1} &=&  -A_1(x') \nonumber\\
{\cal{P}} A_2(x) {\cal{P}}^{-1} &=&  A_2(x') \nonumber\\
{\cal{P}} \psi(x) {\cal{P}}^{-1} &=&  \sigma^1\psi(x')
\label{parity}
\end{eqnarray}
with
\begin{eqnarray}
x &=& (x_0,x_1,x_2) \nonumber\\
x' &=& (x_0,-x_1,x_2) ,
\label{paridad}
\end{eqnarray}
so that the fermion mass term changes sign under parity,

\begin{equation}
{\cal{P}} m \bar \psi (x) \psi (x)  {\cal{P}}^{-1} = - m  \bar \psi (x') \psi
(x') .
\label{roto}
\end{equation}
 The same happens with
time inversion. Analogous results can be derived for any odd dimensional
fermionic fields \cite{Delbourgo}.

Using eqs.(\ref{parity})-(\ref{paridad}) one can easily show that the Dirac
operator Green function changes under parity as follows

\begin{equation}
{\cal{P}} G_{(m)}(x,y)	{\cal{P}}^{-1}= -\sigma^1G_{(-m)}(x',y')\sigma^1
\label{masa}
\end{equation}
This formula (\ref{masa}) is only meaningful for $x \neq y$, where the Green
function is well-defined (we indicate with
the subscript $(m)$ the fact we are considering  the massive Dirac operator
case).
Let us now define the object $ {\cal{J}}^{\mu}_{(m)}(x,y) $,

\begin{equation}
{\cal{J}}^{\mu}_{(m)}(x,y) = tr \gamma^{\mu}G_{(m)}(x,y)
\label{cosa}
\end{equation}
Again, this object is well-defined  whenever $x \neq y$. At $x = y$, precisely
where it gives the ground state current (see eq.(\ref{12})), it has to be
regulated since, appart from regular terms, it has divergent contributions.

 Eq.(\ref{masa}) shows that only for $ m = 0$ ${\cal{J}}^{\mu}_{(m)}(x,y)$
behaves  as a vector
\begin{eqnarray}
{\cal{P}} {\cal{J}}^0_{(0)}(x,y) {\cal{P}}^{-1} &=&  {\cal{J}}^0_{(0)}(x',y')
\nonumber\\
{\cal{P}} {\cal{J}}^1_{(0)}(x,y) {\cal{P}}^{-1} &=&  -{\cal{J}}^1_{(0)}(x',y')
\nonumber\\
{\cal{P}} {\cal{J}}^2_{(0)}(x,y) {\cal{P}}^{-1} &=&   {\cal{J}}^2_{(0)}(x',y')
\label{cosa2}
\end{eqnarray}
(Again, identities (\ref{cosa2}) make sense only for $x \neq y$).
 Then,
as argued in I, parity violating contributions to the ground state
current for {\em massless fermions} cannot arise from regular terms in
 ${\cal{J}}^{\mu}_{(0)}(x,x)$ since these terms should satisfy
 transformation
law (\ref{cosa2}). Only additional regular terms generated by the
 $\zeta$-function prescription in the process of giving
meaning to ${\cal{J}}^{\mu}_{(m)}(x,x)$
can break in this case parity invariance: they are not subject
 to (\ref{cosa2}) but just to respect gauge invariance.
 This is the reason why
we were able to compute in I the {\it complete} parity violating
 contribution
to $j_{\mu}(x)$ for arbitrary $A_{\mu}$: it was not necessary
 to have a complete knowledge of $G_{(0)}(x,y)$ for arbitrary
 $A_{\mu}$ but just the behavior of the additional regular terms
 introduced by the $\zeta$-function prescription.
One should interpret this parity anomalous contribution present
 even for massless fermions as a consequence of regularization
 in odd dimensional space-time.

It is clear from eq.(\ref{cosa}) that the massive case is more involved: appart
from the additional terms generated by the $\zeta$-function regularization,
regular terms which are already present
in  $G_{(m)}(x,x)$ (and a fortiori in ${\cal{J}}^{\mu}_{(m)}(x,x)$) may in
principle contribute to parity violation since they are not constrained to
satisfy vector-like parity transformations. This in turn implies that, in the
massive case, one needs a more
detailed knowledge of $G_{(m)}(x,y)$. For that reason we shall have to  limit
the validity
of ours results to the domain of a perturbative expansion.
With this in mind, we rewrite eq.(\ref{11}) as
\begin{equation}
{j_{\mu}}^{reg}[A]= -tr \left[\gamma_{\mu}\frac{d}{ds}(s
K_{s-1}(x,x;D[A]))\left|\rule[-0.4cm]{0cm}{.4cm}_{s=0}\right.
 \right]
\label{3-1}
\end{equation}
and use the proposition presented in the previous section in the form given by
(\ref{last1})-(\ref{last3}).

We start analysing the regular term in eq.(\ref{last1}) generated by the
$\zeta$-function method,
\begin{equation}
G^{(local)}(x)=
-\int_{|\xi|=1}\frac{d}{ds}C_2(x,\xi;s)\left|\rule[-0.4cm]{0cm}{.4cm}_{s=-1}\right.
\frac{d^2\xi}{(2\pi)^3}.
\label{3-2}
\end{equation}
Using the definition for $C_2$ we have, after differentiating,
\begin{equation}
G^{(local)}(x)= -\frac{i}{2\pi}\int_{|\xi|=1}
\left[ \int_{\Omega} \frac{\ln \lambda}{\lambda}
b_{-3}(x,\xi;\lambda)] d\lambda \right]
\frac{d^2\xi}{(2\pi)^3}.
\label{3-3}
\end{equation}
As the $\xi$-integral is extended to $S_2$, one can see that only terms even in
$\xi$
will give non vanishing contributions to  $G^{(local)}(x)$.
After some algebra one gets from eq.(\ref{20})
\begin{eqnarray}
\lefteqn{G^{(local)}(x)=
-\frac{i}{2\pi}\int_{|\xi|=1} \frac{d^2\xi}{(2\pi)^3}\int_{\Omega} d\lambda
\frac{\ln \lambda}{\lambda (\lambda^2-1)^3}} \nonumber \\
& & \left[ \lambda(e^2A^2-4e^2A_{\mu}A_{\nu}\xi_{\mu}\xi_{\nu} +3m^2 +
2iem\not\!\! A
-8iemA_{\mu}\xi_{\mu} \not\! \xi \right. \nonumber\\
& & + ie\partial_{\mu}A_{\nu} (-4\xi_{\mu}\xi_{\nu}
+\delta_{\mu\nu} + i\epsilon_{\mu\nu\alpha}\gamma_{\alpha})) \nonumber \\
& & \left. +\lambda^3(-e^2A^2 - 2iem\not\!\! A +m^2 -ie\partial_{\mu}A_{\nu}
(\delta_{\mu\nu} + i\epsilon_{\mu\nu\alpha}\gamma_{\alpha}))\right]
\label{3-4}
\end{eqnarray}
We have at this point to choose a curve $\Omega$ in order to perform  the
$\lambda$ integral, this choice
determining the branch for the $\log$ function in eq.(\ref{3-4}). Let us first
consider the curve $\Omega_{(+)}$ depicted
in Fig.\ 1 and call
\begin{equation}
I_{(+)}[p,q]=\int_{\Omega_{(+)}} \frac{\lambda^q
\ln\lambda}{(\lambda^2-1)^p}d\lambda.
\label{Ipq}
\end{equation}
It is easy to see that the integration along the small circle around
the origin vanishes for $q\geq 0$ as its radius goes to zero
while the integrals along
the ray sum up to give
\begin{equation}
I_{(+)}[p,q]=i^q(-1)^p \pi B(\frac{q+1}{2},p-\frac{q+1}{2})
\label{resIpq1}
\end{equation}
where $B(m,n)$ is the $\beta$ function (Euler's integral of the first kind)
\cite{tabla} defined as
\begin{equation}
B(m,n)=\int_0^{\infty}\frac{t^{m-1}}{(1+t)^{m+n}} dt.
\label{beta}
\end{equation}
The $\beta$ functions can be evaluated in terms of $\Gamma$ functions to give
\begin{equation}
I_{(+)}[p,q]=i^q(-1)^p \pi
\frac{\Gamma(\frac{q+1}{2})\Gamma(p-\frac{q+1}{2})}{\Gamma(p)}.
\label{resIpq2}
\end{equation}
Using this result and performing the remaining $\xi$ integral one gets
\begin{equation}
G_{(+)}^{(local)}(x)=-\frac{ie}{8\pi}\epsilon_{\mu\nu\alpha}\partial_{\mu}A_{\nu}
\gamma_{\alpha} + \frac{i}{4\pi}m^2
\label{Gextra}
\end{equation}
where the subscript $(+)$ indicates that we have used the $\Omega_{(+)}$
contour.
The corresponding contribution to the fermionic current
(\ref{3-1}) is
\begin{equation}
{j_{\mu_{(+)}}^{(local)}}= -tr[\gamma_{\mu}G_{(+)}^{(local)}(x)]=
\frac{ie}{4\pi}\epsilon_{\alpha\beta\mu}\partial_{\alpha}A_{\beta}.
\label{resultado+}
\end{equation}

If instead one chooses the curve $\Omega_-$ (see Fig.\ 1) one gets the same
expression but with the opposite sign so we finally have for the regular term
generated by the $\zeta$-function method
\begin{equation}
{j_{\mu_{\pm}}^{(local)}}=\pm\frac{ie}{4\pi}\epsilon_{\alpha\beta\mu}
\partial_{\alpha}A_{\beta}.
\label{jextra}
\end{equation}

Let us first discuss in more detail the origin of the sign ambiguity in
(\ref{jextra}). As discussed in I, $I_{\Omega}[p,q]$ in eq.(\ref{Ipq})
satisfies
\begin{equation}
I_{\Omega_{(-)}}[p,q]=(-1)^{q+1} I_{\Omega_{(+)}}[p,q]
\label{masmenos}
\end{equation}
where $\Omega_{(+)}$ is any curve that avoids a ray of minimal growth of $D[A]$
on the upper half-plane
and $\Omega_{(-)}$ is any other curve that avoids such a ray on the lower
half-plane.
Now, for odd dimensional spaces, only even values for $q$ arise.
In the present $d=3$ case
$q=0,2$. On the other hand, for even dimensional spaces $q=2k+1$.
This is the reason why there are sign ambiguities in computing
anomalies in odd dimensional spaces
(according to the choice of curves $\Omega$), which are not present
for even dimensions.

Another important lesson learnt in this calculation is that this result
is
identical to that corresponding to massless fermions. Indeed,
{\em no new terms} arise from $G_{(\pm)}^{(local)}$ due to the presence of the
mass term since,
as it happens when computing the chiral anomaly \cite{Anna}, the trace in
(\ref{3-1}) makes all the terms
proportional to $m$ and $m^2$ vanish.

We thus see that the contribution to parity violation in
${j_{\mu}}^{reg}[A]$
arising from the regular term
generated by the $\zeta$-function method is independent of whether the fermions
are massless or massive (even if the mass is a regulating mass that
should be taken to zero at the end of the calculation). Its origin can
be traced back to gauge invariant regularization in odd dimensional spaces and
it is present both in the massless and massive case.
\vspace{1cm}

We have now to consider regular terms coming from  the brackets in
eq.(\ref{proposition}), that is, the regular part of $G_{(m)}(x,x)$ remaining
once  the adequated substractions are performed. As stated above, there
was no contribution to parity violation from these terms in the massless case
due to the transformation law
(\ref{masa}) obeyed by	$G_{(0)}(x,x)$: parity was respected at the
classical level so that violations cannot arise
in the process of substracting singular terms to $G_{(0)}(x,x)$.
On the contrary, there may be parity
violating contributions in the case of the massive Green function since already
at the classical level parity is violated by the massive Dirac operator.
Should these contribuitions exist, they will be the product of parity
non-invariance introduced in the Lagrangian
through the fermionic mass term. To see whether this happens, we are
faced to the necessity of making some sort of approximation to compute
$G_{(m)}(x,y)$ at $x=y$ and, from it, the regular part of $G_{(m)}(x,x)$
defined by
$G^{(subst)}(x)$ in eq.(\ref{last2}).

Of course, for $A_{\mu}=0$ the Green function can be computed closely,
\begin{equation}
G_{(m)}(x,y;A_{\mu}=0)= \frac{1}{4\pi}(i\!\!\not\!\partial
-im)\frac{e^{-|m||x-y|}}{|x-y|}.
\label{G0}
\end{equation}
It is then natural to seek for a perturbative expansion in powers of the
coupling constant $e$,
\begin{eqnarray}
\lefteqn{G_{(m)}(x,y;A_{\mu})= G_{(m)}(x,y;A_{\mu}=0)}\nonumber \\
& &  -e \int d^3w
G_{(m)}(x,w;A_{\mu}=0)\not\!\! A(w) G_{(m)}(w,y;A_{\mu}=0)
+ O(e^2)
\label{expansion}
\end{eqnarray}
We will keep only the first order in the perturbative expansion, but a
systematic perturbative calculation can be straightforwardly implemented.
In order to analyse an arbitrary gauge field configuration, we will
make a Taylor expansion of $A_{\mu}(w)$ around
$x$. This leads to a derivative expansion which for dimensional analysis
takes the form of a $\partial/m$ expansion \cite{Babu}.
Up to first derivatives the Green function can be expressed as a Laurent
expansion \mbox{in $z_{\mu}=y_{\mu}-x_{\mu}$ as}
\begin{eqnarray}
\lefteqn{G_{(m)}(x,y;A_{\mu})=\frac{1}{4\pi} \left\{
i\frac{\not\! z}{|z|^3} -im\frac{1}{|z|} -\frac{im^2}{2}\frac{\not\! z}{|z|} +
im|m| \right. } \nonumber \\
& & +eA_{\alpha}(x)\frac{z_{\alpha}\not\! z}{|z|^3}
-emA_{\alpha}(x)\frac{z_{\alpha}}{|z|} \label{G(z)} \\
& & \left. -\frac{ie}{2} \frac{m}{|m|}\epsilon_{\mu\nu\alpha}\partial_{\mu}
A_{\nu}(x) \gamma_{\alpha}
+ \frac{ie}{2} \epsilon_{\mu\nu\alpha}\partial_{\mu} A_{\nu}(x)
\frac{z_{\alpha}}{|z|}
+\frac{e}{2}\partial_{\mu} A_{\nu}(x) \frac{z_{\mu}z_{\nu}\not\! z}{|z|^3}
\right\} +O(z).
\nonumber
\end{eqnarray}

As expected, non-regular terms do appear on the diagonal of the Green function
($z_{\mu} \to 0$). According to Proposition (\ref{proposition}) these terms
should be cancelled by the substractions. Indeed, the explicit
values of these substractions are
\begin{equation}
\int \frac{d^3\xi}{(2\pi)^3}C_1(x,\xi / |\xi|;-1)|\xi|^{-2}e^{i\xi
.(x-y)}=\frac{1}{4\pi}i\frac{\not\! z}{|z|^3},
\label{T0}
\end{equation}
\begin{equation}
\int \frac{d^3\xi}{(2\pi)^3}C_0(x,\xi / |\xi|;-1)|\xi|^{-1}e^{i\xi .(x-y)}
=\frac{1}{4\pi}\left\{	-im\frac{1}{|z|}+eA_{\alpha}(x)\frac{z_{\alpha}\not\!
z}{|z|^3} \right\},
\label{T1}
\end{equation}
and
\begin{eqnarray}
\lefteqn{\int_{|\xi|\geq 1}
\frac{d^3\xi}{(2\pi)^3}C_2(x,\xi;-1)e^{i\xi .(x-y)}=  \frac{1}{4\pi} \left\{
-\frac{im^2}{2}\frac{\not\! z}{|z|}  -emA_{\alpha}(x)\frac{z_{\alpha}}{|z|}
 \right.  }\nonumber \\
& &+ \left.\frac{ie}{2} \frac{m}{|m|}\epsilon_{\mu\nu\alpha}\partial_{\mu}
A_{\nu}(x)
\frac{z_{\alpha}}{|z|}
+\frac{e}{2}\partial_{\mu} A_{\nu}(x) \frac{z_{\mu}z_{\nu}\not\! z}{|z|^3}
\right\}
+O(e^2).
\label{h0resuelto}
\end{eqnarray}
so that using (\ref{proposition}) non-regular terms are exactly cancelled.
In the last expression we used eqs.(\ref{h0MC})-(\ref{M}) where $M(x)=0$; by
$O(e^2)$ we just mean some non-regular terms proportional to $e^2$.
One can
check that each substraction corresponds to non-regular terms homogeneous
of degree -2, -1 and 0 respectively in $z_{\mu}$.

The regular part of $G_{(m)}(x,x)$ remaining after substractions and the
limit $y \to x$  are performed is
\begin{equation}
G^{(subst)}(x)=\frac{1}{4\pi} \{  im|m|  -\frac{ie}{2} \frac{m}{|m|}
\epsilon_{\mu\nu\alpha}\partial_{\mu} A_{\nu}(x) \gamma_{\alpha} \}.
\label{Greg2}
\end{equation}
The corresponding contribution to the current reads
\begin{equation}
{j_{\mu}}^{(subst)}[A]=-tr[\gamma_{\mu}G^{(subst)}(x)]=
\frac{ie}{4\pi}\frac{m}{|m|} \epsilon_{\mu\nu\alpha}\partial_{\nu}
A_{\alpha}(x).
\label{jsubst}
\end{equation}

\vspace{1cm}
The complete expression for the the fermionic
current (up to order $e$ and up to first derivatives of the gauge field)
can now be obtained by adding ${j_{\mu}^{(local)}}$ in eq.(\ref{jextra}) and
${j_{\mu}^{(subst)}}$ in eq.(\ref{jsubst}). The answer
reads
\begin{equation}
{j_{\mu}}^{reg}=\frac{ie}{4\pi}(\pm 1 +\frac{m}{|m|})
\epsilon_{\mu\nu\alpha}\partial_{\nu} A_{\alpha}(x).
\label{jtotal}
\end{equation}

In order to compare this result with those obtained in
previous works, let us first briefly insist on its
domain of validity. Original works showing parity violation
and the consequent emergence of a CS term for massless and massive
fermions were one-loop calculations
performed in a constant $F_{\mu\nu}$ background. For massless fermions
we were able in I, trough
the $\zeta$-function approach, to evaluate the {\em complete} parity violation
contribution for {\em arbitrary} $F_{\mu\nu}$ background. This contribution
reappears
now in the massive case and corresponds to the first term in eq.(\ref{jtotal}).
Again
this contribution corresponds to an {\em exact} result
(not just a one-loop contribution)
valid for {\em arbitrary} $A_{\mu}$.
Now, in the massive case one has also
to consider the  contribution given by $G^{(subst)}(x)$,
i.e.\ the parity violation coming from the fermion mass term. This, we were not
able to evaluate in a closed form due to the impossibility of
calculating the Dirac Green function in an arbitrary
background.
For that reason, we had to appeal to some sort of approximation.
We choosed a perturbative calculation still valid for arbitrary
$F_{\mu\nu}$ and we stopped at the first order in $e$. We also
neglected
$O(\partial^2/m^2)$ in a derivative expansion. Of course our
perturbative expansion can be systematically employed to calculate higher order
contributions to the second term in eq.(\ref{Greg2}).
Although we made the explicit calculations for Abelian gauge fields,
the $\zeta$-function approach can be used with no further modifications
in the non-Abelian case.

Eq.(\ref{jtotal}) is the main result in our paper and we shall devote the next
section to the discussion of its origin and implicancies.
\section{Discussion}
Through the presence of a term proportional to the Levi-Civita pseudo-tensor
in the ground state fermionic current,
eq.(\ref{jtotal}) shows that parity is violated when massive fermions are
considered in three dimensional space time. This in turn produces
a Chern-Simons term in the effective action for the gauge field.

A first important point to remark in formula (\ref{jtotal}) concerns the
existence
of two clearly differenciated sources of parity violation: one is related
with the fact that a mass term neccesary violates P and T in 3 dimensions.
This fact leads to the second term in (\ref{jtotal}). In contrast,
the first term in (\ref{jtotal}) is originated in peculiarities of
regularization in
odd-dimensional
space-times and has nothing to do with the fermion mass. It
is a pure ultraviolet effect (it arises from the necessity of making
regular the singular behavior of the Green function at $x=y$) while the
second term in (\ref{jtotal}) can be interpreted as an infrared
contribution which cannot be naively extended to the $m=0$ case. In fact,
we proved in I that this term is absent if one starts with $m=0$.

Once one gets the $\zeta$-function answer to the parity violation
term in the fermionic current, it is natural to try to understand the
disagreement of the corresponding result with those obtained in
refs.\cite{NS}-\cite{Red}. First, we note that for
$m=0$ there is no disagreement at all, except for the interpretation
of the origin of the anomaly. Indeed, quantitatively, the
answer in \cite{NS}-\cite{Red} and in I was the same. Now, while the
double sign origin was traced back in the former references
to the necessity of introducing regulating mass, it was clear
in the $\zeta$-function approach, where no mass terms is introduced by hand,
that it arises from properties of odd dimensional spaces.

When the fermions are  massive, there are quantitave discrepancies
appart from interpretation differences.
Indeed, the $\zeta$-function answer
contains an additional contribution
not taken into account for example in \cite{Red}. This situation
much resembles what happens in even dimensional spaces concerning gauge
anomalies: also in that case different regularization schemes lead to
two different results for the anomaly in the gauge current conservation
equation. One is known as the covariant anomaly, the other one as the
consistent anomaly (the $\zeta$-function approach leading automatically
to the covariant result; a particular heat-kernel approach leading for
example to the consistent one \cite{Anna}). As explained in \cite{Fuji} it is
on physical
grounds that one should decide to which result attach or, in other words,
which regularization scheme one has to adopt.

Keeping this analogy in mind, we conclude that the particular
physical situation will determine which result
one should use for the parity anomaly. We think that our treatment
has shown, however, the unnaturalness of those proposals in which,
starting with an even
number of fermions and then choosing half  the masses with one sign, half
with the other, parity conservation is achieved: there is a parity violating
effect which is intrinsic to odd dimensions and which cannot be accomodated,
in our opinion, by a clever choice of mass signs. It is there to remain and
it should serve as a guide even when other regularization
prescriptions are adopted.
\section*{Acknowledgements}
F.A.S.\ wishes to acknowledge Nino
Brali\'c whose
systematic criticism on regularization schemes and on the lack of
clarity in Ref.\cite{GMSS} prompted this work.
We would also like to thank Horacio Falomir, Mar\'\i a Amelia Muschietti and
Jorge Solomin
for useful discussions.
R.E.G.S.\ and  G.L.R.\
are
partially supported by CONICET, Argentina. F.A.S.\ is partially
supported by CICBA, Argentina.


\end{document}